\documentclass[
reprint,
superscriptaddress,
amsmath,
amssymb,
aps,
longbibliography,
pra, 
showpacs,
floatfix
]{revtex4-1}

\usepackage{graphicx}
\usepackage{color}

\usepackage{tabularx}
\usepackage{dcolumn}
\newcolumntype{d}[1]{D{.}{.}{#1}}
\usepackage{booktabs}

\usepackage{comment}
\usepackage{enumitem}

\begin{document}

\title{A dynamical magnetic field accompanying the motion of ferroelectric domain walls}

\author{Dominik~M.\ Juraschek}
\email{dominik.juraschek@mat.ethz.ch}
\email{djuraschek@seas.harvard.edu}
\altaffiliation[Current address: ]{Harvard John A. Paulson School of Engineering and Applied Sciences, Harvard University, Cambridge, MA, USA}
\affiliation{Department of Materials, ETH Zurich, Z\"{u}rich, Switzerland}
\author{Quintin N. Meier}
\affiliation{Department of Materials, ETH Zurich, Z\"{u}rich, Switzerland}
\author{Morgan\ Trassin}
\affiliation{Department of Materials, ETH Zurich, Z\"{u}rich, Switzerland}
\author{Susan~E.\ Trolier-McKinstry}
\affiliation{Materials Research Institute, The Pennsylvania State University, University Park, PA, USA}
\author{Christian\ Degen}
\affiliation{Department of Physics, ETH Zurich, Z\"{u}rich, Switzerland}
\author{Nicola~A.\ Spaldin}
\affiliation{Department of Materials, ETH Zurich, Z\"{u}rich, Switzerland}

\date{\today}

\begin{abstract}

The recently proposed dynamical multiferroic effect describes the generation of magnetization from temporally varying electric polarization. Here, we show that the effect can lead to a magnetic field at moving ferroelectric domain walls, where the rearrangement of ions corresponds to a rotation of ferroelectric polarization in time. We develop an expression for the dynamical magnetic field, and calculate the relevant parameters for the example of 90$^\circ$ and 180$^\circ$ domain walls in BaTiO$_3$ using a combination of density functional theory and phenomenological modeling. We find that the magnetic field reaches the order of several $\mu$T at the center of the wall, and we propose two experiments to measure the effect with nitrogen-vacancy center magnetometry.

\end{abstract}

\maketitle



The domain walls that separate different orientations of electric polarization in ferroelectric materials have long been of interest because their motion governs the process of ferroelectric switching in an electric field \cite{Paruch/Giamarchi/Triscone:2007}. Recently, a range of unexpected behaviors have been discovered at domain walls that do not occur in the bulk of the domains, suggesting additional interest in domain walls as functional entities in their own right \cite{Salje:2013}. These include electrical conductivity \cite{Seidel_et_al:2009,Meier_et_al:2012,Sluka_et_al:2013,Mundy2016,smaabraten2018}  or even superconductivity \cite{Aird/Salje:1998} in otherwise insulating systems, ferrielectricity \cite{VanAert_et_al:2011}, as well as magnetoelectricity \cite{Lottermoser_et_al:2004,daraktchiev2010}, strongly anisotropic magnetoresistance \cite{Domingo_et_al:2017} and intriguing dualities between domain walls and the domains themselves \cite{Huang_et_al:2014}.

At the same time, the magnetization caused by the usual motion of electric charges has been revisited over the last years in the context of time-varying ferroelectric polarizations. This newly described {\it dynamical multiferroicity} \cite{juraschek2:2017}, associates a magnetization $\mathbf{M}$ of the form $\mathbf{M}\propto\mathbf{P}\times\partial_t\mathbf{P}$ with a ferroelectric polarization $\mathbf{P}$. A range of existing coupled electric-magnetic phenomena fall within the dynamical multiferroicity framework, and new behaviors, including a phonon Zeeman effect \cite{juraschek2:2017}, exotic quantum criticality \cite{Dunnett2018} and phonon orbital magnetism \cite{Juraschek2018_2} have been proposed. 

Here we discuss the link between these two concepts -- dynamical multiferroicity and ferroelectric domain wall functionality -- by showing theoretically that the motion of ferroelectric domain walls can be accompanied by a dynamical magnetic field. After extending the formalism of dynamical multiferroicity to the case of domain wall motion, we present numerical results for the prototypical ferroelectric barium titanate (BaTiO$_3$), based on first-principles calculations of the polarizations and Born effective charges and on phenomenological modeling using experimental parameters. Finally we discuss the possibility of detecting the dynamical magnetic field experimentally using nitrogen-vacancy center magnetometry.


\begin{figure*}[t]
\centering
\includegraphics[scale=0.1375]{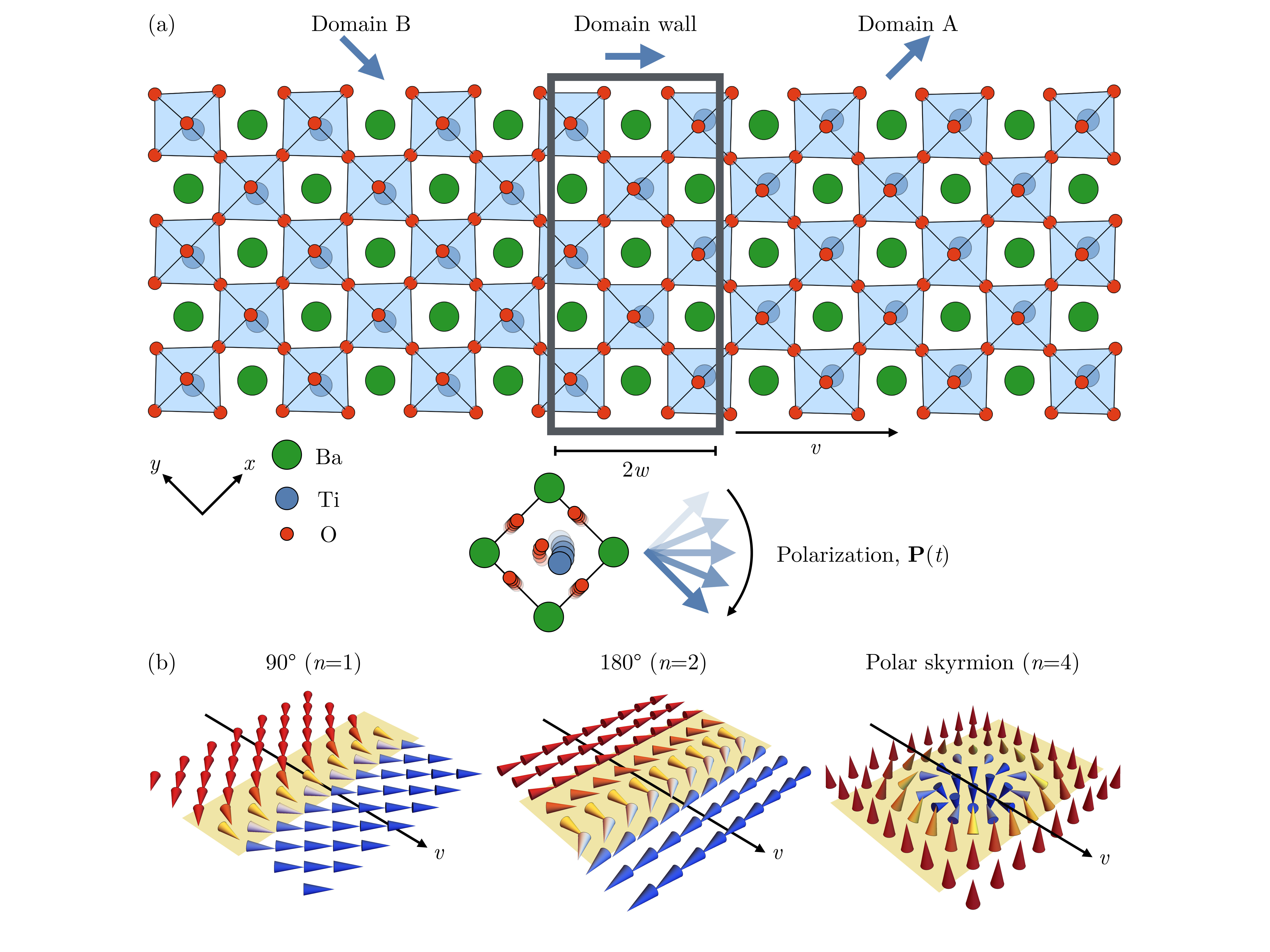}
\caption{
(a) Schematic of a 90$^\circ$ domain wall in the $xy$ plane of BaTiO$_3$ with thickness $2w$ and velocity $v$ separating the two domains A and B. At the bottom, the rearrangement of the titanium and oxygen ions in the moving domain wall is depicted, as well as the change of ferroelectric polarization in time (thick arrows). The displacements of the oxygen and titanium ions are exaggerated for illustration purposes. (b) Illustration of the polarization evolution across 90$^\circ$ and 180$^\circ$ N\'{e}el-type domain walls, and in a polar skyrmion.
}
\label{fig:domainwall}
\end{figure*}


\section*{Theoretical formalism}

We begin by deriving an expression for the dynamical magnetic field at ferroelectric domain walls. Our derivation extends the recently developed microscopic theory for calculating the magnetic moments of optical phonons within the dynamical multiferroicity framework \cite{juraschek2:2017} to the case of moving ionic charges at ferroelectric domain walls. The input parameters in the expression that we obtain can be computed using density functional theory.

The ionic magnetic moment \textbf{m} of a unit cell is given by
\begin{equation}
\mathbf{m} = \sum\limits_i \mathbf{m}_i = \sum\limits_i \gamma_i \mathbf{L}_i, \label{eq:ionicmagmom},
\end{equation}
where $\mathbf{m}_i$ and $\mathbf{L}_i$ are the magnetic moment and the angular momentum arising from the motion of ion $i$, and the sum runs over all ions in the unit cell. $\gamma_i=e\mathbf{Z}^\ast_i/(2\mathcal{M}_i)$ is the gyromagnetic ratio tensor of the ion given by the elementary charge $e$, the Born effective charge tensor $\mathbf{Z}^\ast_i$, and the atomic mass $\mathcal{M}_i$.

The angular momentum at a moving ferroelectric domain wall results from the rearrangement of the atomic positions of the ions in ferroelectric domain A to their respective positions in domain B, see the example of a 90$^\circ$ domain wall in Fig.~\ref{fig:domainwall}a. As the domain wall passes by, the ferroelectric displacement corresponding to domain A, $\mathbf{U}_{i,x}$, reduces to zero, while that corresponding to domain B, $\mathbf{U}_{i,y}$, increases to its bulk value. The angular momentum of ion $i$ can then be written as
\begin{equation}
\mathbf{L}_i = \mathcal{M}_i(\mathbf{U}_{i,x} \times \partial_{t}{\mathbf{U}}_{i,y} + \mathbf{U}_{i,y} \times \partial_{t}{\mathbf{U}}_{i,x}). \label{eq:ionicangularmomentum}
\end{equation}

We can write the time-dependent ferroelectric displacements in terms of a product of the bulk ferroelectric displacement vector $\mathbf{u}_{i,x}$ with a time-dependent dimensionless amplitude $Q_x$, $\mathbf{U}_{i,x}(t) = Q_x (t) \mathbf{u}_{i,x}$. The bulk ferroelectric displacement vector $\mathbf{u}_{i,x} = \mathbf{r}_{i,x} - \mathbf{r}_{i,\mathrm{HS}}$ is given by the difference between the atomic coordinates of the respective ferroelectric structure $\mathbf{r}_{i,x}$, and those of the corresponding high-symmetry structure $\mathbf{r}_{i,\mathrm{HS}}$. Inserting Eq.~(\ref{eq:ionicangularmomentum}) into Eq.~(\ref{eq:ionicmagmom}) we obtain
\begin{eqnarray}
\mathbf{m} & = & \boldsymbol{\gamma} ( \mathbf{Q} \times \partial_{t}{\mathbf{Q}} )_z, \label{eq:domainwallmagmom}
\end{eqnarray}
where $\mathbf{Q}=(Q_x,Q_y,0)$ is the time-dependent amplitude vector and the vector $\boldsymbol{\gamma} = \sum_i \gamma_i \mathcal{M}_i \mathbf{u}_{i,x}\times\mathbf{u}_{i,y}$ given in units of Asm$^2$ contains all atom-specific properties.


\begin{figure*}[t]
\centering
\includegraphics[scale=0.125]{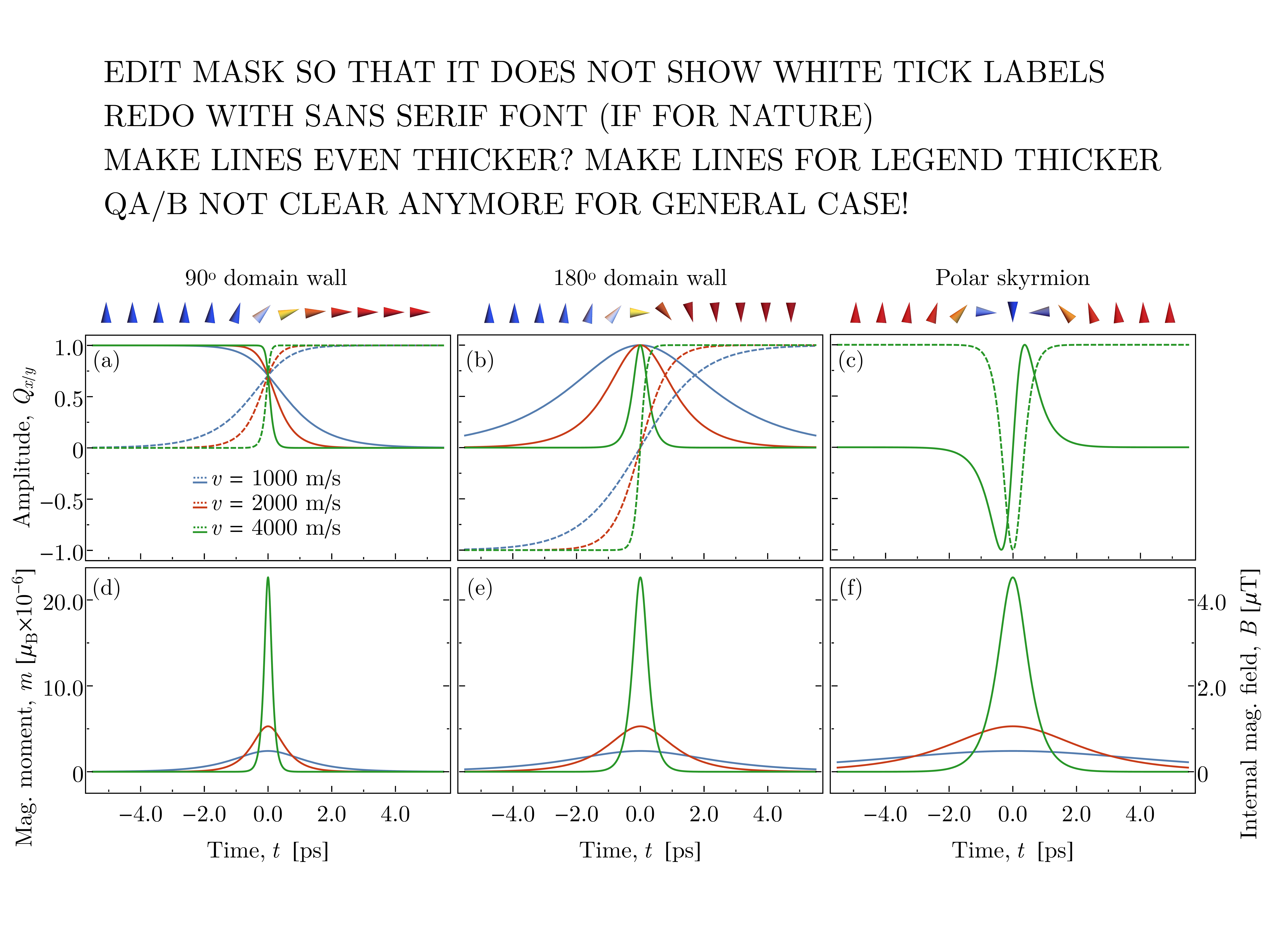} 
\caption{
Time evolution of domain walls. Shown are the dimensionless amplitudes $Q_x$ (solid lines) and $Q_y$ (dashed lines) for the (a) 90$^\circ$ domain wall, (b) 180$^\circ$ domain wall, and (c) polar skyrmion. In (c), the amplitudes are shown only for $v=4000$~m/s for clarity. The resulting magnetic moments $m$ in units of 10$^{-6}$~$\mu_\mathrm{B}$ and internal magnetic fields $B$ in units of $\mu \mathrm{T}$ are shown in (d--f). 
}
\label{fig:magmom}
\end{figure*}

The time evolution of the rotation of polarization in a moving N\'{e}el-type ferroelectric domain wall, in which the ferroelectric polarization rotates within the surface plane, can be described by the sine-Gordon equation with the following solution:
\begin{equation}\label{eq:sinegordonsolution}
\phi(r,t) = \arctan(\mathrm{e}^{\beta(r-vt)/w}),
\end{equation}
see for example Refs.~\cite{Barone1971,Ishibashi1989} and the Supplementary Material \footnote{Supplementary Material}. Here, $\phi$ is the rotation angle, $r$ is the position perpendicular to the domain wall, $v$ is the domain wall velocity, $2w$ is the width of the non-moving domain wall, and the factor $\beta=1/\sqrt{1-v^2/c_0^2}$ describes a Lorentz-like contraction of the moving domain wall that becomes significant for velocities close to the characteristic velocity $c_0$ of the system \cite{collins1979}, which corresponds to the transverse sound velocity \cite{Catalan2012,salje2017}. Without loss of generality we set $r=0$, and with $\phi(0,t)\equiv\phi(t)$ we model the time dependence of $\mathbf{Q}$ as
\begin{equation}\label{eq:amplitudevector}
\mathbf{Q}(t) =
\left(
\begin{array}{c}
Q_x(t) \\
Q_y(t)
\end{array}
\right)
=
\left(
\begin{array}{c}
\cos(n\phi(t)) \\
\sin(n\phi(t))
\end{array}
\right),
\end{equation}
where $n$ determines the amount of polarization rotation between the domains ($n=1$ for a 90$^\circ$ domain wall).

Inserting Eqs.~(\ref{eq:sinegordonsolution}) and (\ref{eq:amplitudevector}) into Eq.~(\ref{eq:domainwallmagmom}), the ionic magnetic moment per unit cell accompanying the motion of the domain wall reduces to
\begin{equation}\label{eq:evaluatedmagmom}
\mathbf{m}(t) = \boldsymbol{\gamma} n \partial_{t}{\phi}(t),
\end{equation}
The internal magnetic field \textbf{B} created by the ionic magnetic moment of the moving domain wall is then given by $\textbf{B} = \mu_0\textbf{m}/V$, where $\mu_0$ is the vacuum permeability and $V$ the volume of the unit cell. 


\section*{Numerical results}

In this section, we estimate the magnitude of the dynamical magnetic field according to Eq.~\ref{eq:evaluatedmagmom} starting with the example of a 90$^\circ$ domain wall in BaTiO$_3$, see Fig.~\ref{fig:domainwall}. The fundamental input parameters to Eq.~(\ref{eq:evaluatedmagmom}) are the Born effective charge tensors $\mathbf{Z}^\ast_i$, the ferroelectric displacement vectors $\mathbf{u}_{i,x/y}$, and the volume of the unit cell $V$, which we calculate from first-principles, as well as the domain wall thickness $2w$ and the domain wall velocity $v$ which we take from experimental literature. (For details of the first-principles calculations, see the Methods section.) Experimental values for $2w$ and $v$ vary strongly throughout the literature, and we therefore estimate them within realistic boundaries. Reported 90$^\circ$ domain wall thicknesses of BaTiO$_3$ range between 2 to 25~nm \cite{Hlinka2006}. Domain wall velocities have been reported up to several times 10$^3$~m/s for 90$^\circ$ domain wall wedges in BaTiO$_3$ \cite{Stadler1964,Faran2010}, as well as several times 10$^3$~m/s for other domain walls in related ferroelectrics \cite{Meng2015}. The ultimate barrier for $v$ is the transverse sound velocity of the material, which in BaTiO$_3$ is $4.4\times10^3$~m/s \cite{Merz1956}.


\begin{figure*}[t]
\centering
\includegraphics[scale=0.11]{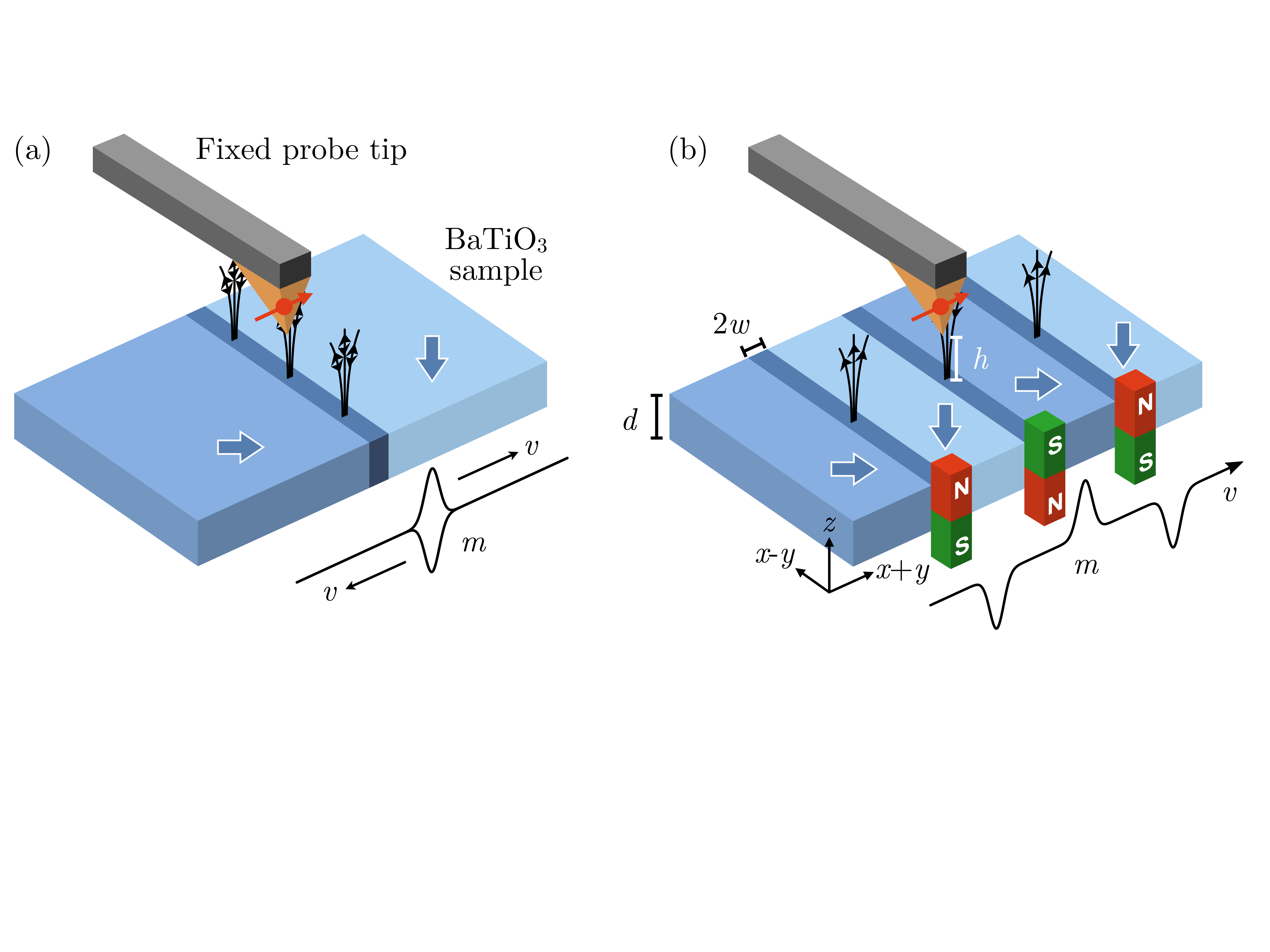}
\caption{
Proposed experiments to measure the dynamical magnetic field induced by the motion of a ferroelectric domain wall. (a) A time-dependent electric field induces a back-and-forth motion (shivering) of a single domain wall, generating an oscillating magnetic field. (b) A time-dependent electric field in combination with a sawtooth potential provided by the substrate induces unidirectional motion of multiple domain walls as in Ref.~\cite{Whyte2015}, generating a magnetic field pulse train with alternating sign.
}
\label{fig:experiment}
\end{figure*}

We show the time evolution of the two amplitudes $U_x$ and $U_y$ in Fig.~\ref{fig:magmom}a, and the ionic magnetic moment of the unit cell and its corresponding magnetic field for a domain wall with thickness $2w=2$~nm and different values of the domain wall velocity $v$ in Fig.~\ref{fig:magmom}d. The quicker the rearrangement, meaning the thinner and faster the domain wall, the larger is the peak magnetic field. For the largest velocity of $v=4000$~m/s that we show here, the ionic magnetic moment per unit cell reaches $m=24.7\times10^{-6}$~$\mu_\mathrm{B}$, which corresponds to an internal magnetic field of $B=4.5$~$\mu$T.

We now extend our calculations to other types of domain walls: to 180$^\circ$ N\'{e}el-type domain walls as were predicted in lead titanate (PbTiO$_3$) \cite{Lee2009} and recently observed in lead zirconium titanate (PZT) films \cite{DeLuca2017,Cherifi-Hertel2017}, as well as to recently predicted polar skyrmions, in which electric dipole moments form a spiral structure \cite{Nahas2015}. We assume pure N\'{e}el character for both cases, and consequently the 180$^\circ$ case can straightforwardly be treated as an extension of the 90$^\circ$ case with $n=2$ in Eq.~\ref{eq:amplitudevector}. The center of a moving N\'{e}el-type polar skyrmion can be treated as a 360$^\circ$ domain wall between domains of the same orientation of polarization with $n=4$; in this case the polarization rotates within a plane perpendicular to the surface and to the domain wall, which causes the magnetic field to lie in the surface plane.

We show the time evolution of the amplitudes $U_x$ and $U_y$ for the two cases in Figs.~\ref{fig:magmom}b and c, and the ionic magnetic moment of the unit cell and its corresponding internal magnetic field in Figs.~\ref{fig:magmom}e and f. Here, we use a thickness of $2w=4$~nm for the 180$^\circ$ domain wall and $2w=8$~nm for the polar skyrmion. By construction, for these wall widths the ionic magnetic moment per unit cell and the internal magnetic field yield the same values as the 90$^\circ$ case, however, with double and four times the full width at half maximum duration of the peak, respectively.


\section*{Possible experimental realization}

NV center defects in diamond have emerged during the past decade as an ultrasensitive detection tool for nanoscale magnetic fields \cite{Degen2008,Maze2008,Taylor2008,Casola2018}. NV centers carry a single electron spin, whose response to changes in the local magnetic field can be observed through their paramagnetic resonance transition \cite{Degen2008}. In the field of ferroelectrics, NV center scanning probes have been used to probe magnetic domain walls in multiferroic bismuth ferrite (BiFeO$_3$) \cite{Gross2017}. Here, we propose their use in probing the magnetic field at a moving ferroelectric domain wall, exploiting recent improvements in control of ferroelectric domain wall motion, down to the single domain-wall level \cite{McGilly2015}.

We show schematic setups of two complementary experiments that could be used to detect the dynamical magnetic field induced by the motion of a ferroelectric domain wall in Fig.~\ref{fig:experiment}. In our proposed experiments, a fixed probe using an NV center in a diamond tip is placed above the surface of a BaTiO$_3$ sample. In the first setup, a time-dependent electric field, suitably shielded from the probe tip, induces back-and-forth motion (shivering) of the domain wall, see Fig~\ref{fig:experiment}a. The magnetic stray field produced by the moving wall will change sign depending on the direction of the domain wall motion, resulting in an oscillatory magnetic signal at the probe tip that is tuned to the resonance frequency of the NV center. In the second setup, motion of multiple domain walls past the tip is induced, see Fig~\ref{fig:experiment}b. The magnetic moments accompanying each domain wall act as a magnetic pulse train that changes sign as successive domain walls pass the probe tip. If the sizes of the domains are roughly equal, the domain wall velocity is chosen such that the oscillatory magnetic signal caused by the pulse train is tuned to the resonance frequency of the NV center. Unidirectional motion of domain walls has been demonstrated by applying a time-dependent electric field in combination with a sawtooth potential provided by the substrate, which prevents backwards motion over the sawtooth potential edge \cite{Whyte2015}.

We estimate the magnetic stray field at the position of a probe tip located at height $h>w$ above the surface of BaTiO$_3$ as follows. The effective driving field (or Rabi field) $B_1$ is given by $B_1 = \gamma n 2\pi \mu_0  f_0 /V$, where $\gamma=|\boldsymbol{\gamma}|$ and $f_0$ is the NV center spin's resonance frequency. (For a derivation, see the Methods section.) Our calculations yield a value of $\gamma=4.8\times 10^{-41}$ Ams$^2$. For an NV center near zero bias field, $f_0 \approx 2.87~\mathrm{GHz}$, giving a Rabi field of $B_1 \approx 17~\mathrm{nT}$ for the 90$^\circ$ domain wall configuration of Fig.~\ref{fig:experiment}a ($n=1$). For a 180$^\circ$ domain wall, the Rabi field accordingly doubles, $B_1 \approx 34~\mathrm{nT}$. These values lie above achievable sensitivities at room temperature and well above sensitivities projected to be achievable at liquid nitrogen temperatures \cite{Joas2017}.


\section*{Methods}


\subsection*{First-principles calculations}

We calculate the Born effective charges, the ferroelectric displacements, and the unit cell volume from first-principles using the density functional theory formalism as implemented in the Vienna ab-initio simulation package (VASP) \cite{kresse:1996,kresse2:1996}. We use the VASP projector augmented wave (PAW) pseudopotentials with valence electron configurations Ba 5s$^2$5p$^6$6s$^2$, Ti 3d$^3$4s$^1$, and O 2s$^2$2p$^4$ and converge the Hellmann-Feynman forces to 0.1~meV/\AA. For the 5-atom unit cell, we use a plane-wave energy cut-off of 850~eV, and an 8$\times$8$\times$8 gamma-centered $k$-point mesh to sample the Brillouin zone. For the exchange-correlation functional, we choose the Perdew-Burke-Ernzerhof revised for solids (PBEsol) form of the generalized gradient approximation (GGA) \cite{csonka:2009}. The lattice constants of our fully relaxed tetragonal structure (space group $P4mm$) of $a=3.983$~\AA{} and $c=4.031$~\AA{} with a unit cell volume of $V=63.9$~\AA$^3$, as well as the calculated ferroelectric polarization of $22.9$~$\mu$C/cm$^2$ match reasonably well with experimental values \cite{Choi2004}. For a list of the calculated Born effective charges see the Supplementary Material.


\subsection*{Stray field estimate}

The vertical magnetic stray field $B_z^\mathrm{s}$ appearing at height $h$ above a thin ($w\ll h$), extended domain wall located (see Fig.~\ref{fig:experiment}a) is given by
\begin{align}
B_z^\mathrm{s}(r) = \frac{\mu_0 M_\text{2d} h}{2\pi(r^2+h^2)} \ ,
\end{align}
where $r$ is the relative position of the domain wall with respect to the sensor, 
$M_\text{2d}$ is the two-dimensional moment density of the domain wall, and where the material is assumed to be thick ($d\gg w,h$). The magnetic moment density is given by
\begin{align}
M_\text{2d} = \frac{1}{V}\int_{-\infty}^{\infty} \mathrm{d}r m_z(r) = \frac{1}{V}\int_{-\infty}^{\infty} \mathrm{d}t v \gamma n \dot\phi(t) = \frac{\gamma}{V}v n \pi.
\end{align}

To detect the magnetic stray field, we repeatedly move the domain wall back and forth to create an oscillatory field $B_z^\mathrm{s}(t)$ at the location of the NV spin sensor.  When tuned to the spin resonance frequency $f_0$, the oscillatory field induces a Rabi rotation of the spin providing an experimentally detectable signal.
The Rabi field $B_1$ is given by the Fourier component of $B_z^\mathrm{s}(t)$ at frequency $f_0$,
\begin{align}
B_1 = \frac{4}{T} \int_{-T/4}^{T/4} \mathrm{d}t \cos(2\pi t/T) B_z^\mathrm{s}(t),
\end{align}
where $T = 1/f_0$ is the Larmor period of the spin.  Because the Larmor period ($\sim 0.1-1~\mathrm{ns}$) is typically much longer than the time taken for the domain wall to pass ($\sim 1~\mathrm{ps}$), $B_z^\mathrm{s}(t)$ is non-zero only for a very short period of time around $t=0$, and the integral can be approximated by
\begin{eqnarray}
B_1
  & \approx & \frac{4}{T} \int_{-\infty}^{\infty} \mathrm{d}t B_z^\mathrm{s}(t)
	 =  \frac{4}{T} \frac{\mu_0 m h}{2\pi} \int_{-\infty}^{\infty} \frac{\mathrm{d}t}{v^2 t^2+h^2} \nonumber\\
	& = & \frac{\gamma}{V} n 2\pi \mu_0  f_0.
\end{eqnarray}
Note that the expected signal is independent of domain wall width $w$, speed $v$ and sensor height $h$ as long as the geometric assumptions about the setup (a thick sample with $h\gg w$) are satisfied and $vT\gg h$.


\section*{Conclusion}

In summary, we have identified that moving ferroelectric domain walls have a magnetization resulting from the dynamical multiferroic effect. We predict the magnetic moment accompanying the domain wall motion to reach up to 25 micro $\mu_\mathrm{B}$ per unit cell, corresponding to the order of several $\mu$T. The Rabi field generated by the domain wall of up to 34~nT lies within the range of experimentally achievable sensitivities of NV center magnetometry.

In addition to the 90$^\circ$ and 180$^\circ$ N\'{e}el-type domain walls and polar skyrmions studied in this work, our proposed mechanism is generally applicable. We expect, for example, the 71$^\circ$ and 109$^\circ$ domain walls in bismuth ferrite (BiFeO$_3$) or PZT to exhibit strong effects because of their large Born effective charges and domain wall velocities. The mechanism is also valid for Bloch-type domain walls, in which the ferroelectric polarization rotates within a plane perpendicular to the surface, and in this case produces magnetizations lying within the surface plane. For pure Ising-type domain walls, in which the polarization reduces to zero at the center of the wall with no perpendicular components, the angular momentum at the domain wall and therefore the effect is zero.

We hope that our proposal sparks experimental efforts to realize the mechanism, adding yet another manipulable degree of freedom to the functionality of domain walls. Experimental success in measuring our proposed phenomenon may result in improved characterization of ferroelectric domain wall motion \cite{Catalan2012,Sharma2017}, and in detecting possible motion of polar skyrmions and polar vortices \cite{Nahas2015,Yadav2016}.


\begin{acknowledgments}
We are grateful to Pietro Gambardella for useful discussions. This work was supported by the ETH Zurich. Calculations were performed at the Swiss National Supercomputing Centre (CSCS) supported by the project IDs s624, p504.
C.L.D. acknowledges funding by the Swiss National Science Foundation under Grant No. 200020-175600 and the NCCR QSIT, and by the European Commission through Grant No. 820394 “ASTERIQS”.
\end{acknowledgments}



%


\onecolumngrid
\clearpage

\setcounter{page}{1}

\begin{center}
\textbf{\large A dynamical magnetic field accompanying the motion of ferroelectric domain walls: Supplementary Material}\\[0.4cm]
Dominik~\,M.~\,Juraschek,$^{1,\ast}$ Quintin N. Meier,$^1$ Morgan\ Trassin,$^1$ \\
Susane\ E.\ Trolier-McKinstry,$^2$ Christian\ Degen,$^3$ and Nicola\ A.\ Spaldin$^1$\\[0.15cm]
\affiliation{}
$^1${\itshape{\small Department of Materials, ETH Zurich, Z\"{u}rich, Switzerland}}\\
$^2${\itshape{\small Materials Research Institute, The Pennsylvania State University, University Park, PA, USA}}\\
$^3${\itshape{\small Department of Physics, ETH Zurich, Z\"{u}rich, Switzerland}}\\
\end{center}

\setcounter{equation}{0}
\setcounter{figure}{0}
\setcounter{table}{0}
\makeatletter
\renewcommand{\theequation}{S\arabic{equation}}
\renewcommand{\thefigure}{S\arabic{figure}}
\renewcommand{\thetable}{S\arabic{table}}

\section*{Time evolution of N\'{e}el- and Bloch-type ferroelectric domain walls}

In the following we review the derivation of the domain wall motion based on Refs.~\cite{Barone1971,Ishibashi1989}, rewriting it in the notation used in this work. The Lagrangian for a ferroelectric domain wall separating domains of different orientation of polarization lying in the $xy$ plane can be written as
\begin{eqnarray}\label{SUPPeq:lagrangian}
\mathcal{L} & = & \frac{1}{2} \mathcal{M}_x (\partial_t U_x)^2 - \frac{1}{2} S_x (\partial_r U_x)^2 - \frac{1}{2} A_x U_x^2 - \frac{1}{4} B_x U_x^4 \nonumber\\
& + & \frac{1}{2} \mathcal{M}_y (\partial_t U_y)^2 - \frac{1}{2} S_y (\partial_r U_y)^2 - \frac{1}{2} A_y U_y^2 - \frac{1}{4} B_y U_y^4 \nonumber\\
& - & C_{xy} U_x^2 U_y^2,
\end{eqnarray}
where $U_{x/y}$ is the amplitude of the ferroelectric displacement along direction $x/y$, $\mathcal{M}_{x/y}$ is the effective mass of the ferroelectric distortion mode, $S_{x/y}$ are the gradient energies and $A_{x/y}$, $B_{x/y}$ and $C_{xy}$ are the coefficients of the harmonic, quartic anharmonic, and coupling terms. The coordinate $r$ denotes the position perpendicular to the domain wall in the $xy$ plane of polarization. For a Bloch-type domain wall, we would require components perpendicular to the plane of polarization, $z\perp x,y$; for a N\'{e}el-type domain wall, we can express the change of ferroelectric displacement as simple rotation in the $xy$ plane:
\begin{equation}\label{SUPPeq:amplitude}
\mathbf{U} = 
\left(
\begin{array}{c}
U_x (t) \\
U_y (t)
\end{array}
\right)
=
U_0
\left(
\begin{array}{c}
\cos(\phi(r,t)) \\
\sin(\phi(r,t))
\end{array}
\right),
\end{equation}
where $U_0$ is the amplitude of the bulk ferroelectric displacement. ($\mathbf{U}=U_0\mathbf{Q}$ in Eq.~\ref{eq:amplitudevector} in the main text with $n=1$.) Inserting this into the Lagrangian~(\ref{SUPPeq:lagrangian}), together with $\mathcal{M}_x = \mathcal{M}_y = \mathcal{M}$, $S_x = S_y \equiv S$, $A_x=A_y\equiv A$, $B_x=B_y\equiv B$, $B_x=B_y\equiv B$, and $C_{xy} \equiv C$ we obtain
\begin{eqnarray}\label{SUPPeq:lagrangianphi}
\mathcal{L} = \frac{1}{2} \mathcal{M} U_0^2 (\partial_t \phi)^2 - \frac{1}{2} S U_0^2 (\partial_r \phi)^2 - \frac{1}{2} A U_0^2 - \frac{1}{4} B U_0^4 - C U_0^4 \frac{1}{8}(1-\cos(4\phi)).
\end{eqnarray}
The Euler-Lagrange equations for the Lagrangian~(\ref{SUPPeq:lagrangianphi}) yield after some rearrangements
\begin{equation}\label{SUPPeq:eulerlagrange}
\partial_t^2 \phi - c_0^2 \partial_r \phi + \kappa^2 \sin(4\phi) = 0,
\end{equation}
where $\kappa^2=C U_0^2 / (2\mathcal{M})$ and $c_0=\sqrt{S/\mathcal{M}}$ is the characteristic velocity. A substitution $4\phi\rightarrow\theta$ and a transformation to a moving frame $r\rightarrow\xi=r-vt$, where $v$ is the constant domain wall velocity yields
\begin{equation}\label{SUPPeq:eulerlagrangexi}
\partial_\xi^2 \theta - \alpha^2 \sin(\theta) = 0,
\end{equation}
where $\alpha^2 = \kappa^2/(c_0^2(1-v^2/c_0^2))$. The solution to this equation is known for a 360$^\circ$ rotation of $\theta$ (corresponding to a 90$^\circ$ rotation of $\phi$), see for example Ref.~\cite{Ishibashi1989}:
\begin{eqnarray}
\theta(\xi) & = & 4 \arctan\left(\mathrm{e}^{\xi \alpha}\right) \label{SUPPeq:solution}\\
\Rightarrow \phi(r,t) & = & \frac{1}{4}\theta(\xi(r,t)) = \arctan\left(\exp\left(\frac{1}{w}\frac{r-vt}{\sqrt{1-\frac{v^2}{c_0^2}}}\right)\right), \label{SUPPeq:solutionfull}
\end{eqnarray}
where $w=c_0^2/\kappa^2=S/(2CU_0^2)$ and $2w$ is the width of the domain wall. Eq.~(\ref{SUPPeq:solutionfull}) is the expression (\ref{eq:sinegordonsolution}) given in the main text. $n\phi$ with $n=1,2,4$ then corresponds to 90$^\circ$, 180$^\circ$, and 360$^\circ$ rotations of the ferroelectric polarization.

\pagebreak


\section*{Born effective charges}

\begin{table}[h]
\centering
\bgroup
\def\arraystretch{1.25}
\caption{
Calculated diagonal Born effective charges $Z^\ast_{i,aa}$ in units of the elementary charge ($a$ denoting spatial coordinates $x$, $y$, and $z$), and ferroelectric displacements $u_{i,x}$ in picometers for polarization along the $x$ direction.
}
\begin{tabularx}{0.25\textwidth}{lcccc}
\hline\hline
\multicolumn{1}{l}{Atom~~} & \multicolumn{1}{c}{$Z^\ast_{i,xx}$} & \multicolumn{1}{c}{$Z^\ast_{i,yy}$} & \multicolumn{1}{c}{$Z^\ast_{i,zz}$} & \multicolumn{1}{c}{$u_{i,x}$} \\
\hline
Ba    &  2.8 &  2.7 &  2.7 &  4.2   \\
Ti    &  6.6 &  7.6 &  7.6 &  8.9   \\
O(1)  & -5.3 & -2.1 & -2.1 & -3.4   \\
O(2)  & -2.0 & -6.0 & -2.1 & -0.8   \\
O(3)  & -2.0 & -2.1 & -6.0 & -0.8   \\
\hline\hline
\end{tabularx}
\label{tab:ferroelectricdata}
\egroup
\end{table}

\end{document}